\documentclass[12pt,showpacs,preprintnumbers,amsmath,aps,amssymb]{revtex4}
\usepackage{graphicx}
\usepackage{dcolumn}
\usepackage{bm}
\usepackage{color}
\begin{document}

\baselineskip 18pt

\title{Anisotropic fractional cosmology: K-essence theory}

\author{J. Socorro$^1$ }
\email{socorro@fisica.ugto.mx}
\author{J. Juan Rosales$^2$}
\email{rosales@ugto.mx}
\author{L. Toledo Sesma$^3$}
\email{ltoledos@ipn.mx}
\affiliation{$^1$ Departamento de F\'{\i}sica de la Divisi\'on de Ciencias e Ingenier\'ias de la Universidad de Guanajuato-Campus Le\'on,\\
 C.P. 37150,  Guanajuato, M\'exico.\\
$^2$ Departamento de Ingenier\'ia El\'ectrica, DICIS. Universidad de Guanajuato,\\
Comunidad de Palo Blanco, C.P. 36885, Salamanca Guanajuato,
M\'exico.\\
$^3$ Instituto Polit\'ecnico Nacional, Unidad Profesional Interdisciplinaria de Ingenier\'ia Campus Hidalgo, Carretera Pachuca - Actopan Kil\'ometro 1+500, 42162.\\
San Agust\'in Tlaxiaca, Hgo. M\'exico}

\begin{abstract}
In the particular configuration of the scalar field K-essence in the
Wheeler-DeWitt quantum equation, for some age in the Bianchi type I
anisotropic cosmological model, a fractional differential equation
for the scalar field arises naturally. The order of the fractional
differential equation is $\beta=\frac{2\alpha}{2\alpha - 1}$. This
fractional equation belongs to different intervals, depending on the
value of the barotropic parameter; when $\omega_{X} \in [0,1]$, the
order belongs to the interval $1\leq \beta \leq 2$, and when
$\omega_{X}\in[-1,0)$, the order belongs to the interval $0< \beta
\leq 1$. In the quantum scheme, we introduce the factor ordering
problem in the variables $(\Omega,\phi)$ and its corresponding
momenta $(\Pi_\Omega, \Pi_\phi)$, obtaining a linear fractional
differential equation with variable coefficients in the scalar field
equation, then the solution is found using a fractional power series
expansion. The corresponding quantum solutions are also given. We
found the classical solution in the usual gauge N obtained in the
Hamiltonian formalism and without a gauge. In the last case, the
general solution is presented in a  transformed time $T(\tau)$,
however in the dust era we found a closed solution in the gauge time
$\tau$.

Keywords:  Fractional derivative, Fractional Quantum Cosmology;
K-essence formalism; Classical and Quantum exact solutions.
\end{abstract}

\maketitle

\section{Introduction}
Fractional cosmology is a new line of research born approximately
twenty years ago based on fractional calculus (FC).  The FC is a
non-local natural generalization to the arbitrary order of
derivatives and integrals.  Non-local effects occur in space and
time. In the time domain, a non-local description becomes manifest
as a memory effect, and  in the space domain, it manifests as
non-homogeneous similarity structures
 \cite{Podlubny, Uchaikin, Herman}. During the last decades, FC has been the subject of intense  theoretical and applied research, almost in all areas of the sciences and engineering,  from the point of view of the classical and quantum systems
 \cite{Ca, Wyss, Westerlund, r-s, Magin, Tarasov, Duarte, Laskin, Paulo2020, Paulo2021a, Paulo2023},  recently new studies on FC have been made \cite{Ortigueira2022a, Ortigueira2022b, Ortigueira2023a, Ortigueira2023b}.
 This is because, the FC describes more accurately the complex physical systems and at the same time, investigates  more about simple dynamical systems \cite{Rosales1,Rosales2}. The general relativity could not be the exception,
 in \cite{Rami1, Rami-2012, Rami-2013, Rami2015, Rami2016a, Rami2016b, Rami2017a, Rami2017b, Paulo2021, Paulo2022a, Paulo2022b}
 the importance of FC and its potential applications in cosmology was introduced. In \cite{Rami2} the FRW universe  was presented in the context of the variational principle of fractional action. In this new cosmological formulation,  the accelerated expansion of the universe can be attributed to the fractional dissipative force without the need
 to introduce any kind of matter or scalar fields, similar results are obtained in \cite{Aspeitia1, Aspeitia2}.
In \cite{Rami-2012}, the concept of fractional action cosmology was
applied to massive gravity, where fractional graviton masses are
introduced.

Unlike the previously described formalism to obtain fractional
cosmology, in \cite{Socorro1} it is mentioned that by quantifying
different epochs of the K-essence theory, a fractional
Wheeler-DeWitt equation in the scalar field component is naturally
obtained.  Recently, such an equation was solved for some epochs in
the FRW model and communicated in \cite{Socorro2}.
 In this work we present the continuation of our previous investigation, in this case, we will analyze the Bianchi type I,  which is the anisotropic generalization of the flat FRW cosmological  model. In the quantum scheme, we introduce the factor ordering problem in the variables $(\Omega,\phi)$ and its corresponding momenta ($\Pi_\Omega, \Pi_\phi$), obtaining a fractional differential equation with variable coefficients in the
scalar field equation. The solution is found using a fractional
series expansion \cite{El-Ajou,Rida}, generalizing our previous work
\cite{Socorro2}.

This paper is organized in such a way: in Section 1, we give a brief
review of fractional calculus and the main ideas of the K-essence
formalism; in Section 2, we construct the Lagrangian and Hamiltonian
densities for the anisotropic Bianchi type I cosmological model,
considering a barotropic perfect fluid for the scale field in the
variable $X$. We found the classical solution in the usual gauge N
obtained in the Hamiltonian formalism, and without a gauge. In the
last case, the general solution is presented in a transformed time
$T(\tau)$, however in the dust era we found a closed solution in the
gauge time $\tau$; in Section 3, the quantization of the model for
any era in our universe is done and present particular scenarios,
too. In this section, we introduce the factor ordering in both
variables; finally, in Section 4, the conclusions are given.

\section{Brief review on fractional calculus and K-essence theory}
\subsection{Brief review on fractional calculus}
In the theory of fractional calculus, there are some definitions of
fractional derivatives; Rieman-Liouville, Caputo, Caputo-Fabrizio,
Atangana-Baleanu, to name a few, each with its advantages and
disadvantages \cite{Ortigueira, Tarasov1, Sales}. In this work, we
use the Caputo fractional derivative of order $\gamma$, defined by
using the Riemann-Liouville fractional integral \cite{Podlubny}
\begin{equation}
I^\gamma f(t) = \frac{1}{\Gamma(\gamma)} \int_0^t \frac{f(\tau)}{( t
- \tau)^{1 - \gamma}} d\tau, \quad \gamma > 0, \label{FI}
\end{equation}
recovering ordinary integral, when $\gamma \to 1$. The Caputo
fractional derivative of order $\gamma \geq 0$ of a
 function $f(t)$, then, is defined as the fractional order integral (\ref{FI}) of the integer order derivative
\begin{equation}
^C_0D^{\gamma}_{t} f(t) = I^{(n - \gamma)}\, _{0}D_t^n f(t) =
\frac{1}{\Gamma(n - \gamma)}\int_0^t \frac{f^{(n)}(\tau)}{( t -
\tau)^{\gamma - n +1}} d\tau, \label{C1}
\end{equation}
with $n-1 <\gamma \leq n \in \mathbb{N} = {1,2, ...}$, and $\gamma
\in \mathbb{R}$ is the order of the fractional derivative and
$f^{(n)}$ are the ordinary integer derivatives, and $\Gamma(x) =
\int_0^\infty e^{-t} t^{x - 1} dt$, is the gamma function. The
Caputo derivative satisfies the following relations
\begin{eqnarray}
^C_0D^\gamma ( f(t) + g(t)] &=& ^C_0D^\gamma f(t) + {^C_0}D^\gamma  g(t). \label{C2}\\
^C_0D^\gamma c &=& 0,\qquad {\rm where}\,\, c\,\, {\rm is\,\, a\,\,
constant}.  \label{C3}
\end{eqnarray}

The Laplace transform of the function $f(t)$ defined in the ordinary
case is given by
\begin{equation}
    \mathbb{L} [f(t)] = \int_0^\infty f(t) e^{-st} dt \equiv F(s), \label{Laplace1}
\end{equation}
then, the Laplace transform of the Caputo fractional derivative
(\ref{C1}) has the form
\begin{equation}
\mathbb{L}[^C_0D^\gamma f(t) ] = s^\gamma F(s) - \sum_{k=0}^{n-1}
s^{\gamma - k - 1} f^{(k)}(0), \label{C4}
\end{equation}
where $f^{(k)}$ is the ordinary derivative. Another definition which
will be used is the Mittag-Leffler function
\cite{Erdely,Haubold,Trifce},
\begin{equation}
\mathbb{E}_{\chi, \sigma}(z) = \sum_{n=0}^\infty \frac{z^n}{\Gamma(n
\chi  + \sigma)}\qquad (\chi,\,\,\, \sigma > 0), \label{C5}
\end{equation}
for $\sigma = 1$, we have one parameter Mittag-Leffler function
\begin{equation}
\mathbb{E}_{\chi}(z) = \mathbb{E}_{\chi,1}(z) = \sum_{n=0}^\infty
\frac{z^n}{\Gamma(n\chi + 1)} \qquad (\chi > 0). \label{C6}
\end{equation}
Another special cases are in \cite{Erdely,Haubold}
\begin{eqnarray}
\mathbb{E}_{1}(\pm z) &=& e^{\pm z},\qquad \mathbb{E}_{2}(z) = \cosh{\sqrt{z}},\qquad \mathbb{E}_{2,1}(-z^2) = \cos{z}, \nonumber\\
\mathbb{E}_{2,2}(z^2) &=& \frac{\sinh{z}}{z},\qquad
\mathbb{E}_{2,2}(-z^2) = \frac{\sin{z}}{z}.
\end{eqnarray}
Laplace transform (\ref{Laplace1}) of the Mittage-Leffler function
is given by the formula
\begin{equation}
\int_0^\infty e^{-st} t^{\chi m + \sigma -
1}\mathbb{E}_{\chi,\sigma}^{(m)} (\pm at^\chi) dt = \frac{ m!\,
s^{\chi - \sigma}}{(s^\chi \mp a )^{m+1}}. \label{C7}
\end{equation}
Consequently, the inverse Laplace transform is
\begin{equation}
\mathbb{L}^{-1}\Big[  \frac{ m!\, s^{\chi - \sigma}}{(s^\chi \mp a
)^{m+1}}  \Big] =  t^{\chi m + \sigma -
1}\mathbb{E}_{\chi,\sigma}^{(m)} (\pm at^\chi).
\end{equation}
This expression will be very useful to obtain analytical solutions
of fractional differential equations using the Laplace transform.

\subsection{K-essence fractional in the Bianchi I scenario.}
One of the fundamental problems of cosmology is to find an
explanation consistent with experiments for the accelerated
expansion of the universe. Many proposals to tackle this task
suggest modifying the general relativity theory. A recent proposal
suggests unifying the description of dark matter, dark energy and
inflation, employing a scalar field with a nonstandard kinetic term,
known as K-essence theory. Usually, the action of the K-essence
models
\cite{Espinoza,dePutter:2007ny,Chiba:2009nh,Bose:2008ew,Arroja:2010wy,Garcia:2012aa}
can be written as
\begin{equation}
S = \int\,d^{4}x\sqrt{-g}\biggl(\frac{1}{2}\mathcal{R} +
f(\phi)\mathcal{G}(X) + {\cal L}_{matter}\biggr), \label{ac1}
\end{equation}
$g$ being the determinant of the metric, $\mathcal{R}$ the scalar curvature, $f(\phi)$ an arbitrary function of the dimensionless scalar field $\phi$, $X = -\frac{1}{2}g^{\mu \nu}\nabla_{\mu}\phi\nabla_{\nu}\phi$ the canonical kinetic energy and ${\cal L}_{matter}$ is the corresponding Lagrangian density of ordinary matter. 
So, performing the variation of the action \eqref{ac1} with respect
to the metric $g_{\mu\nu}$ and $X$, the field equations are obtained
\begin{eqnarray}
 G_{\mu \nu} - f(\phi)\biggl[\mathcal{G}_{X}\nabla_{\mu}\phi\nabla_{\nu}\phi + \mathcal{G}g_{\mu \nu}\biggr] & =&T_{\mu \nu},
\label{matt_eq} \\
f(\phi)\biggl[\mathcal{G}_{X}\nabla^{\mu}\nabla_{\mu}\phi +
\mathcal{G}_{XX}\nabla_{\mu}X\nabla^{\mu}\phi\biggr] +
\frac{df}{d\phi}\biggl[\mathcal{G} - 2X\mathcal{G}_{X}\biggr] & =&
0, \label{kg-eq}
\end{eqnarray}
where, we have assumed that $8\pi G = 1$ and a subscript $X$ denotes
differentiation with respect to $X$. K-essence was originally
proposed as a model for inflation; and then, as a model for dark
energy, along with explorations of unifying dark energy and dark
matter \cite{Bilic:2001cg,Bento:2002ps}.

Last set of field equations (\ref{matt_eq}) and (\ref{kg-eq}) are
the results of considering the scalar field $X(\phi)$ as part of the
matter content, \textit{i.e.} $\mathcal{L}_{X,\phi} =
f(\phi)\mathcal{G}(X)$, with the corresponding energy-momentum
tensor

\begin{equation}
\mathcal{T}_{\mu \nu}^{(\phi)} =
f(\phi)\left[\mathcal{G}_{X}\nabla_{\mu}\phi\nabla_{\nu}\phi +
\mathcal{G}(X)g_{\mu \nu}\right]. \label{tenmat}
\end{equation}
Also, considering the energy-momentum tensor of a barotropic perfect
fluid,
\begin{equation}
T_{\mu \nu}^{(\phi)} = (\rho_\phi + P_\phi)u_{\mu}u_{\nu} + P_\phi
g_{\mu \nu}, \label{tenbar}
\end{equation}
with $u_{\mu}$ being the four-velocity satisfying the relation
$u_{\mu}u^{\mu} = -1$, $\rho_\phi$ the energy density and $P_\phi$
the pressure of the fluid. To simplify, we are going to consider a
comoving perfect fluid, whose pressure and energy density
corresponding to the energy moment tensor of the field X are
\begin{equation}
P_\phi (X) = f(\phi)\mathcal{G}, \qquad \rho_\phi (X) =
f(\phi)\left[2X\mathcal{G}_{X} - \mathcal{G}\right], \label{charten}
\end{equation}
thus the barotropic parameter $\omega_{X} = \frac{P_\phi
(X)}{\rho_\phi (X)}$ for the equivalent fluid is

\begin{equation}
\omega_{X} = \frac{\mathcal{G}}{2X\mathcal{G}_{X} - \mathcal{G}}.
\label{barpar}
\end{equation}
Notice that the case of a constant barotropic index $\omega_{X}$
(with the exception $\omega_{X} = 0$) can be obtained by the
$\mathcal{G}$ function
\begin{equation}
\mathcal{G} = X^{\frac{1 + \omega_{X}}{2\omega_{X}}}. \label{expfor}
\end{equation}
At this point we can choose
\begin{equation}
\mathcal{G} = X^{\alpha},\qquad \alpha=\frac{1 +
\omega_{X}}{2\omega_{X}} \qquad \to \qquad \omega_{X} =
\frac{1}{2\alpha - 1}. \label{expge}
\end{equation}
 With this, we can write the states in the evolution of the universe resumed in the table \ref{staevo}.
\begin{table}[!htbp]
  \centering
  \begin{tabular}{|c|c|c|c|}
\hline
  $\omega_{X}$ & $\alpha$ & $\mathcal{G}(X)$ & State of evolution \\
\hline\hline
1 & 1 & $X$ & Stiff matter \\
\hline
$\frac{1}{3}$  & 2 & $X^{2}$ & Radiation \\
\hline
$\rightarrow 0$  & $\rightarrow \infty $ & $X^{m}$, \quad $m\rightarrow\infty$ & Dust like \\
\hline
$-1$ & $0$ & $1$, \quad $f(\phi) = \Lambda = cte$ & Inflation \\
\hline
$-\frac{1}{3}$ & -1 & $\frac{1}{X}$ & Inflation like \\
\hline
$-\frac{2}{3}$  & $-\frac{1}{4}$ & $\frac{1}{\sqrt[4]{X}}$ & Inflation like \\
\hline
  \end{tabular}
  \caption{States of the universe's evolution according to the barotropic parameter $\omega_{X}$.}
  \label{staevo}
\end{table}

\vspace{1cm}

We are interested in the four-dimensional fractional cosmology in
the scenario of k-essence within the anisotropic background,
precisely the Bianchi type I, whose metric has the line element
$g_{\alpha\beta}$, which can be read as
\begin{equation}
ds^2  = -N^2(t)\,dt^2 + A^2(t)\,dx^2 + B^2(t)\,dy^2 + C^2(t)\,dz^2,
\label{metric-ii}
\end{equation}
where $N(t)$ is the lapse function, the functions $A(t)$, $B(t)$ and
$C(t)$ are the corresponding scale factors in the $(x,y,z)$
directions, respectively. Moreover, in the Misner's parametrization,
the radii for this anisotropic background have the explicit form
\begin{equation}
A=e^{\Omega+\beta_+ +\sqrt{3}\beta_-}, \qquad B=e^{\Omega+\beta_+
-\sqrt{3}\beta_-}, \qquad C=e^{\Omega-2\beta_+}, \label{radBI}
\end{equation}
where the functions in the radii are dependent on time, $\Omega =
\Omega(t)$ and $\beta_{\pm} = \beta_{\pm}(t)$. In this point, we
notice that the line element (\ref{metric-ii}), in the time
$d\tau=Ndt$ reads as
\begin{equation}
ds^2=-d\tau^2 +e^{2(\Omega(\tau) + \beta_+(\tau) +
\sqrt{3}\beta_-(\tau))}\,dx^2 + e^{2(\Omega(\tau) + \beta_+(\tau) -
\sqrt{3}\beta_-(\tau))}\,dy^2 + e^{2\Omega(\tau)-4\beta_+(\tau)}\,
dz^2, \label{metric-iii}
\end{equation}
and employing the form of the functional ${\cal G}=X^\alpha$, and
the following quantities
\begin{eqnarray}
&& \prime=\frac{d}{d\tau}=\frac{d}{Ndt},\qquad   g_{\tau
\tau}=g^{\tau \tau} = -1,\qquad {\cal G}_X=\alpha
X^{\alpha-1}, \qquad{\cal G}_{XX}=\alpha(\alpha-1) X^{\alpha-2},\nonumber\\
&&  \phi^{,\mu}_{\,;\mu}=g^{\mu\nu}\left(\phi_{,\mu \nu}-\Gamma_{\mu
\nu}^\theta \phi_{,\theta} \right)=-\left(\phi^{\prime
\prime}+3\Omega^\prime \phi^\prime \right),\qquad
X_{;\mu}\phi^{,\mu}=g^{\mu \nu}\phi_{,\nu}X_{;\mu}=-X^\prime
\phi^\prime,\nonumber\\
&& X=\frac{1}{2}\left(\phi^{\prime} \right)^2, \qquad
\left(\phi^{\prime} \right)^2=2X, \qquad  X^\prime=\phi^\prime
\phi^{\prime \prime}, \qquad  \phi^{\prime
\prime}=\frac{X^\prime}{\phi^\prime},\nonumber
\end{eqnarray}
then the Equation~(\ref{kg-eq}) is written as ,
\begin{equation}
 \alpha X^{\alpha-1}\left(\phi^{\prime \prime}+3 \Omega^\prime \phi^\prime \right)+ \alpha(\alpha-1) X^{\alpha-2}X^\prime \phi ^\prime
 +(2\alpha-1) X^\alpha \frac{d}{d\phi}Ln f=0,
\label{Sfe-m}
\end{equation}
which can be transformed into
\begin{equation}
\frac{d}{d\tau}\left( Ln X + \frac{6\Omega}{2\alpha-1}+Ln
f^{\frac{1}{\alpha}}\right)=0, \label{master}
\end{equation}
and in turn integrated, resulting
\begin{equation}
\int f^{\frac{1}{2\alpha}}(\phi)\,d\phi=\sqrt{2\lambda} \int
e^{-\frac{3\Omega(\tau)}{2\alpha-1}}d\tau, \label{mm}
\end{equation}
where $\lambda$ is an integration constant and has the same sign as
$f(\phi)$. In the gauge $N=24 e^{\frac{3\Omega}{2\alpha-1}}$ the
right side is
\begin{equation}
\int f^{\frac{1}{2\alpha}}(\phi)\,d\phi=24\sqrt{2\lambda}\,\,
(t-t_i), \label{mmp}
\end{equation}
where $t_i$ is the initial time for the $\alpha$ scenario in the
universe. At this point, we can introduce some structure for the
function $f(\phi)$ and solve the integral.

When we consider the particular mathematical structure for the
function $f(\phi)=p \phi^m$ or $f(\phi)=p e^{m\phi}$ with $p$ and
$m$ constants, the classical solutions for the field $\phi$ in
quadratures are
\begin{equation}
\phi(\tau)=\phi(\tau_i) + \left\{
    \begin{tabular}{ll}
    $\left[\frac{(2\alpha+m)}{2\alpha}{p^{-\frac{1}{2\alpha}}\sqrt{2\lambda }}\int e^{-\frac{3\Omega(\tau)}{2\alpha-1}}d\tau\right]^{\frac{2\alpha}{2\alpha+m}},$&\quad $f(\phi)=p\phi^m, \quad m\not=-2\alpha$\\
$    Exp\left[ {p^{-\frac{1}{2\alpha}}\sqrt{2\lambda }} \int
e^{-\frac{3\Omega(\tau)}{2\alpha-1}}d\tau\right],$
         &\quad $f(\phi)=p\phi^{-2\alpha}, \quad m=-2\alpha$  \\
          $\frac{2\alpha}{m}\,Ln\left[\frac{m}{2\alpha} {p^{-\frac{1}{2\alpha}}\sqrt{2\lambda }}\int e^{-\frac{3\Omega(\tau)}{2\alpha-1}}d\tau\right],$ & \quad $f(\phi)=pe^{m\phi},\quad m\not=0,$\\
         ${p^{-\frac{1}{2\alpha}}\sqrt{2\lambda }}\int e^{-\frac{3\Omega(\tau)}{2\alpha-1}}d\tau$ & \quad $f(\phi)=p, \quad m=0$.
    \end{tabular}
    \right. \label{scalar-field}
\end{equation}
The complete solution to the scalar field $\phi$ depends strongly on
the  mathematical structure of the scale factor $\Omega(\tau)$ in
the $\alpha$ scenario in our universe.
In the gauge $N=24 e^{\frac{3\Omega}{2\alpha-1}}$, these solutions
are
\begin{equation}
\phi(t)=\phi(t_i) + \left\{
    \begin{tabular}{ll}
    $\left[\frac{12(2\alpha+m)}{\alpha}{{p^{-\frac{1}{2\alpha}}\sqrt{2\lambda }}}  \left(t-t_i\right)\right]^{\frac{2\alpha}{2\alpha+m}},$&\quad $f(\phi)=p\phi^m, \quad m\not=-2\alpha$\\
$    Exp\left[ 24{{p^{-\frac{1}{2\alpha}}\sqrt{2\lambda }}} \left( t
- t_i \right)\right],$
         &\quad $f(\phi)=p\phi^{-2\alpha}, \quad m=-2\alpha$  \\
          $\frac{2\alpha}{m}\,Ln\left[\frac{12m}{\alpha} {{p^{-\frac{1}{2\alpha}}\sqrt{2\lambda }}}\left(t-t_i \right)\right],$ & \quad $f(\phi)=p\,e^{m\phi},\quad m\not=0,$\\
         $24{{p^{-\frac{1}{2\alpha}}\sqrt{2\lambda }}}\left(t-t_i \right)$ & \quad $f(\phi)=p, \quad m=0$,
    \end{tabular}
    \right. \label{phi-t}
\end{equation}
where $t_i$ and $\phi(t_i)$ are the initial time and the scalar
field in this time for the $\alpha$ scenario in the universe. In
what follows, we do the calculations to obtain the scale factor in
some cases.

\section{Lagrange and Hamilton formalism}

Introducing the line element (\ref{metric-ii}) of the anisotropic
Bianchi type I cosmological model into the Lagrangian (\ref{ac1}),
we have

\begin{equation}
\mathcal{L}_{I}= e^{3\Omega}\left\{6\frac{\dot\Omega^2}{N} -
6\frac{\dot\beta_+^2}{N} - 6\frac{\dot\beta_-^2}{N} - f(\phi)
\left(\frac{1}{2}\right)^{\alpha}\left(\dot\phi\right)^{2\alpha}
N^{-2\alpha+1}\right\}. \label{lagra-i}
\end{equation}
 Using the standard definition of the momenta
$\Pi_{q^{\mu}}=\frac{\partial\mathcal{L}}{\partial\dot{q}^{\mu}}$,
where $q^{\mu}$ are the coordinate fields $q^{\mu}=(\Omega,
\beta_\pm, \phi)$, we obtain the momenta associated with each field

\begin{alignat}{2}
\Pi_{\Omega} &= \frac{12}{N}e^{3\Omega}\dot \Omega, &\qquad   \dot{\Omega} &=e^{-3\Omega} \frac{N\Pi_{\Omega}}{12}, \nonumber\\
\Pi_{\pm} &= -\frac{12}{N}e^{3\Omega}\dot \beta_\pm, &\qquad  \dot{\beta_\pm} &= -e^{-3\Omega} \frac{N\Pi_{\pm}}{12},,\nonumber\\
 \Pi_{\phi} &=
-
f(\phi)\left(\frac{1}{2}\right)^\alpha\frac{2\alpha}{N^{2\alpha-1}}\,e^{3\Omega}{\dot
\phi}^{2\alpha -1}, &\qquad \dot \phi
&=-Ne^{-\frac{3\Omega}{2\alpha-1}}\left[\frac{2^\alpha}{2\alpha
}\frac{\Pi_\phi}{f(\phi)}\right]^{\frac{1}{2\alpha -1}}, \label{ph}
\end{alignat}
and introducing them into the Lagrangian density, we obtain the
canonical Lagrangian as $\mathcal{L}_{canonical} =\Pi_{q^\mu} \dot
q^\mu - N\mathcal{H}=\Pi_{q^\mu} \dot q^\mu - H$. When we perform
the variation of this canonical Lagrangian with respect to $N$,
$\frac{\delta\mathcal{L}_{canonical}}{\delta N} =0$, we obtain the
constraint $\mathcal{H}=0$. In our model, this is the only
constraint corresponding to the Hamiltonian density, which is weakly
zero. So, the Hamiltonian is
\begin{equation}
H=\frac{N}{24}e^{-\frac{3}{2\alpha-1}\Omega} \left\{
e^{-\frac{6(\alpha-1)}{2\alpha-1}\Omega}\left[\Pi_\Omega^2 - \Pi_+^2
- \Pi_-^2\right] -
\frac{12(2\alpha-1)}{\alpha}\left(\frac{2^{\alpha-1}}{\alpha
f(\phi)}
\right)^{\frac{1}{2\alpha-1}}\,\Pi_\phi^{\frac{2\alpha}{2\alpha-1}}\right\}.
\label{hami}
\end{equation}

\subsection{Exact solution in the gauge $N=24e^{\frac{3}{2\alpha - 1} \Omega}$.}
Using the Hamilton equations for the momenta
 $\dot{\Pi}_{\mu}=-\frac{\partial H}{\partial q^\mu}$ and coordinates $\dot{q}^\mu =\frac{\partial H}{\partial \Pi_{\mu}}$,
we have
\begin{eqnarray}
\dot{\Omega}&=& 2 e^{-\frac{6(\alpha-1)}{2\alpha-1}\Omega} \Pi_\Omega\label{a}\\
\dot{\beta_+}&=& -2 e^{-\frac{6(\alpha-1)}{2\alpha-1}\Omega} \Pi_+, \label{bb1}\\
\dot{\beta_-}&=& -2 e^{-\frac{6(\alpha-1)}{2\alpha-1}\Omega} \Pi_-, \label{bb2}\\
\dot \phi &=&  -24\left(\frac{2^{\alpha-1}}{\alpha f(\phi)}
\right)^{\frac{1}{2\alpha-1}}\,\Pi_\phi^{\frac{1}{2\alpha-1}} , \label{f-phi}\\
\dot{\Pi}_\Omega &=& \frac{6(\alpha-1)}{2\alpha-1}
e^{-\frac{6(\alpha-1)}{2\alpha-1}\Omega}\left[\Pi_\Omega^2 - \Pi_+^2
- \Pi_-^2 \right] , \label{pa}\\
 \dot{\Pi}_\pm &=& 0,\qquad \qquad \qquad \Rightarrow \qquad
\Pi_\pm=p_\pm=constant, \label{p}\\
 \dot{\Pi}_{\phi}&=&-\frac{12}{\alpha}\left(\frac{2^{\alpha-1}}{\alpha} \right)^{\frac{1}{2\alpha-1}} \Pi_\phi^{\frac{2\alpha}{2\alpha-1}}f^{-\frac{2\alpha}{2\alpha-1}}\frac{\dot f}{\dot \phi},  \label{pphi}
\end{eqnarray}
solving the equation (\ref{pphi}) using \eqref{f-phi}, we have
$\Pi_\phi=p_\phi f^{\frac{1}{2\alpha}}$, with $p_\phi$ an
integration constant. With this result and taking into account the
equation \eqref{f-phi}, we get
\begin{equation}
\int f^{\frac{1}{2\alpha}}\,d\phi= -24\left(\frac{p_\phi
2^{\alpha-1}}{\alpha} \right)^{\frac{1}{2\alpha -1}} \left(t-t_i
\right),
\end{equation}
that is, similar to (\ref{mmp}), previously obtained, when was
solved a Klein-Gordon like equation, directly. Using the Hamiltonian
constraint and the solution to equation \eqref{pphi} found
previously, we have
\begin{equation}
e^{-\frac{6(\alpha-1)}{2\alpha-1}\Omega}\left[\Pi_\Omega^2 - \Pi_+^2
- \Pi_-^2\right] =
\frac{12(2\alpha-1)}{\alpha}\left(\frac{p_\phi^{2\alpha}\,2^{\alpha-1}}{\alpha}
\right)^{\frac{1}{2\alpha-1}},
\end{equation}
then, the solution for the momenta becomes
\begin{equation}
\Pi_\Omega = \eta_\alpha\,t +p_0,
\end{equation} where the constant
$\eta_\alpha=\frac{72(\alpha-1)}{\alpha}
\left(\frac{p_\phi^{2\alpha}\,2^{\alpha-1}}{\alpha}
\right)^{\frac{1}{2\alpha-1}} $, and $p_0$ are constants of
integration,
 that introducing in the equation for $\Omega$, we get the equation for the $\Omega$ function
\begin{equation}
\frac{d\Omega}{dt}=2 e^{-\frac{6(\alpha-1)}{2\alpha-1}\Omega}
\left(\eta_\alpha t + p_0\right),  \qquad
e^{\frac{6(\alpha-1)}{2\alpha-1}\Omega}=\frac{6(\alpha-1)}{2\alpha-1}\left(
\eta_\alpha t^2 + 2p_{0}t + p_1\right), \label{sol-omega}
\end{equation}
whose solution becomes
\begin{equation}
\Omega=\frac{2\alpha-1}{6(\alpha-1)}\,
\ln{\left[\frac{6(\alpha-1)}{2\alpha-1}\left(\eta_\alpha t^2+2p_0t
+p_1 \right) \right]},
\end{equation}
and the solution for the scalar field is given by the equations
\eqref{phi-t}. The solutions for the anisotropic function
$\beta_\pm$ are given by

\begin{equation}
\beta_\pm(t)=b_\pm - \frac{(2\alpha-1)p_{\pm}}{24( \alpha -
1)\eta_{\alpha}\sqrt{p_{0}^{2} - \eta_{\alpha} p_{1}}}\, \ln{\left[
\frac{\eta_\alpha t+p_0-\sqrt{p_0^2-\eta_\alpha p_1}} {\eta_\alpha
t+p_0+\sqrt{p_0^2-\eta_\alpha p_1}}\right]},
\end{equation}

\begin{equation}
\beta_{\pm} = b_{\pm} - \left(\frac{2\alpha - 1}{24(\alpha -
1)}\right)\frac{p_{\pm}}{\eta_{\alpha}\sqrt{\lambda_\alpha}}\,\ln{\left(\frac{\Sigma_{-}(t)}{\Sigma_{+}(t)}\right)},
\end{equation}
where

\begin{equation*}
\Sigma_{\pm}(t) = \eta_{\alpha}t + p_{0} \pm \sqrt{\lambda_\alpha},
\qquad \text{and} \qquad \lambda_\alpha = p_{0}^{2} -
\eta_{\alpha}p_{1} > 0.
\end{equation*}
According to the last expressions, the radii associated with the
Bianchi I have the following behaviour

\begin{subequations}
\begin{align}
A(t) & = e^{\Omega + \beta_{+} + \sqrt{3}\,\beta_{-}} = A_{0}\Biggl[\frac{6(\alpha - 1)}{2\alpha - 1}\Biggl(\frac{\Sigma_{+}(t)}{\Sigma_{-}(t)}\Biggr)^{\frac{p_{+} + \sqrt{3}\,p_{-}}{4\eta_{\alpha}\sqrt{\lambda_\alpha}}}(\eta_{\alpha}t^{2} + 2p_{0}t + p_{1})\Biggr]^{\frac{2\alpha - 1}{6(\alpha - 1)}}, \\
B(t) & = e^{\Omega + \beta_{+} - \sqrt{3}\,\beta_{-}} = B_{0}\Biggl[\frac{6(\alpha - 1)}{2\alpha - 1}\Biggl(\frac{\Sigma_{+}(t)}{\Sigma_{-}(t)}\Biggr)^{\frac{p_{+} - \sqrt{3}\,p_{-}}{4\eta_{\alpha}\sqrt{\lambda_\alpha}}}(\eta_{\alpha}t^{2} + 2p_{0}t + p_{1})\Biggr]^{\frac{2\alpha - 1}{6(\alpha - 1)}}, \\
C(t) & = e^{\Omega - 2\beta_{+}} = C_{0}\Biggl[\frac{6(\alpha
-1)}{2\alpha -
1}\Biggl(\frac{\Sigma_{-}(t)}{\Sigma_{+}(t)}\Biggl)^{\frac{p_{+}}{2\eta_{\alpha}\sqrt{\lambda_\alpha}}}(\eta_{\alpha}t^{2}
+ 2p_{0}t + p_{1})\Biggr]^{\frac{2\alpha - 1}{6(\alpha - 1}},
\end{align}
\label{radBIs}
\end{subequations}
and the volume of this universe $V(t) = ABC = e^{3\Omega}$ being,
\begin{equation}
V(t) = V_{0}\Biggl[\frac{6(\alpha - 1)}{2\alpha -
1}(\eta_{\alpha}t^{2} + 2p_{0}t + p_{1})\Biggr]^{\frac{2\alpha -
1}{2(\alpha - 1)}}.
\end{equation}

\begin{figure}[!htbp]
    \centering
    \includegraphics[width=0.4\textwidth]{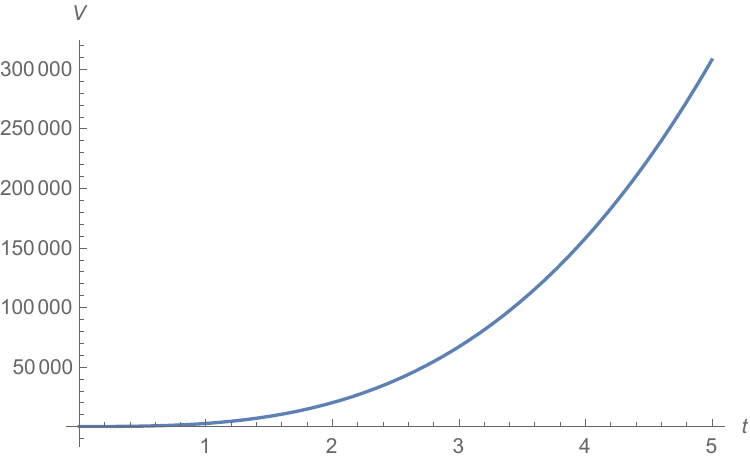}
    \includegraphics[width=0.4\textwidth]{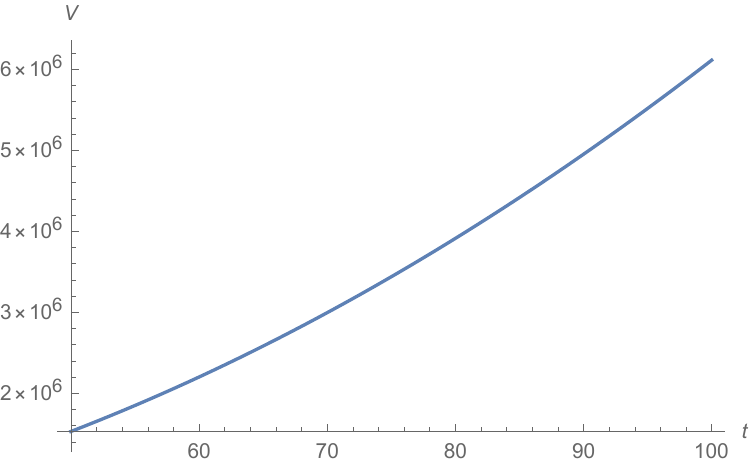}
    \caption{Volume of the universe into the radiation and Dust age, respectively, according to the table \ref{staevo}, we choose $p_\phi=2$ and $p_1=1$.}
    \label{plot1}
\end{figure}
Where we have graphed on different time scales in each scenario, in
both cases the volume is increasing, as is shown in figure
\ref{plot1}.

\subsection{Exact solution without gauge N in the time $\tau$}
For this case, the  Hamilton procedure is not adequate, then we
shall use the Hamilton-Jacobi procedure in order to find the
solutions for the remaining minisuperspace variables, which arises
by making the identification  $\frac{\partial
S(\Omega,\beta_\pm,\phi)}{\partial q_\mu}=\Pi_\mu$ in the
Hamiltonian constraint (\ref{hami}), $H = 0$, taking $
S(\Omega,\beta_\pm,\phi)=S_\Omega(\Omega)+
S_+(\beta_+)+S_-(\beta_-)+S_\phi(\phi)$ which results in
\begin{equation}
e^{-\frac{6(\alpha-1)}{2\alpha-1}\Omega}\left[\left(\frac{d
S_\Omega}{d \Omega}\right )^2 - \left (\frac{d S_+}{d \beta_+}\right
)^2 - \left(\frac{d S_-}{d \beta_-}\right )^2\right] -A_\alpha
\left(\frac{1}{f(\phi)} \right)^{\frac{1}{2\alpha-1}}
\,\left(\frac{d S_\phi}{d
\phi}\right)^{\frac{2\alpha}{2\alpha-1}}=0. \label{hamilton-jacobi}
\end{equation}
Separating this equation, we have
\begin{equation}
e^{-\frac{6(\alpha-1)}{2\alpha-1}\Omega}\left[\left(\frac{d
S_\Omega}{d \Omega}\right )^2 - \left (\frac{d S_+}{d \beta_+}\right
)^2 - \left(\frac{d S_-}{d \beta_-}\right )^2\right] =A_\alpha
\left(\frac{1}{f(\phi)} \right)^{\frac{1}{2\alpha-1}}
\,\left(\frac{d S_\phi}{d
\phi}\right)^{\frac{2\alpha}{2\alpha-1}}=\ell_\phi^2.
\label{separate-phi}
\end{equation}
with $\ell_\phi$ a separation constant. The solution in the variable
$\phi$ is,
\begin{equation}
 \Pi_\phi=\frac{d S_\phi}{d
\phi}=\left[\frac{\ell_\phi^2}{A_\alpha}\right]^{\frac{2\alpha-1}{2\alpha}}
f^{\frac{1}{2\alpha}}(\phi) =p_\phi
f^{\frac{1}{2\alpha}}(\phi),\label{phi-hj}
\end{equation}
where $S(\phi)=p_\phi \int f^{\frac{1}{2\alpha}}(\phi) \,d\phi$,
obtaining similar results in the Hamilton procedure.

The specific values of the constants are
$p_\phi=\left[\frac{\ell_\phi^2}{A_\alpha}\right]^{\frac{2\alpha-1}{2\alpha}}$
and $A_\alpha=
\frac{12(2\alpha-1)}{\alpha}\left(\frac{2^{\alpha-1}}{\alpha }
\right)^{\frac{1}{2\alpha-1}}$, in terms of the $\alpha$ parameter.


The other equations are read as
\begin{eqnarray}
\left(\frac{dS_\Omega}{d\Omega} \right)^2 &=& \ell_+^2 +
\ell_-^2+\ell_\phi^2 \,
e^{\frac{6(\alpha-1}{2\alpha-1}}, \label{omega1}\\
\left(\frac{dS_+}{d\beta_+} \right)^2 &=& \ell_+^2,\qquad S_+=s_+ \pm \ell_+ \, \beta_+, \label{beta+1}\\
\left(\frac{dS_-}{d\beta_-} \right)^2 &=& \ell_-^2,\qquad S_-=s_-
\pm \ell_- \, \beta_-, \label{beta-1}
\end{eqnarray}
where $\ell_i^2$ are separation constants and $s_\pm$ integration
constants. On the other side, recalling the expressions for the
momenta we can obtain solutions for equations \eqref{omega1},
\eqref{beta+1} and \eqref{beta-1} in quadrature; for the variable
$\Omega$ and for $\alpha \not= 1$,

\subsection{Case for $\alpha \not=1$}
In this  particular case, we have
\begin{equation}
d \tau= 12 \frac{e^{3\Omega}\,d\Omega}{\sqrt{\ell^2 + \ell_\phi^2 \,
e^{\frac{6(\alpha-1)}{2\alpha-1}\Omega}}}, \qquad
\ell^2=\ell_+^2+\ell_-^2, \label{omega2}
\end{equation}
and for the anisotropic variables,
\begin{equation}
\Delta \beta_\pm= \mp \frac{\ell_\pm}{12} \int e^{-3\Omega(\tau)}\,
d\tau. \label{ani}
\end{equation}
For solving equation \eqref{omega2}, we employ the transformation in
the time variable $d\tau=e^{\frac{3}{2\alpha-1}\Omega} dT$ and
${\cal
U}=\ell^2+\ell_\phi^2\,e^{\frac{6(\alpha-1)}{2\alpha-1}\Omega}$, so,
$d{\cal U}=\ell_\phi^2 \frac{6(\alpha-1)}{2\alpha-1}
e^{\frac{6(\alpha-1)}{2\alpha-1}\Omega} d\Omega$, resulting
$$dT=12\frac{e^{\frac{6(\alpha-1)}{2\alpha-1}\Omega} d\Omega}{\sqrt{{\cal
U}}}=\frac{2(2\alpha-1)}{(\alpha-1)\ell_\phi^2} \frac{d{\cal
U}}{\sqrt{{\cal U}}},$$ and the solution is
$$T-T_0 =\frac{4(2\alpha-1)}{(\alpha-1)\ell_\phi^2}\left(\sqrt{{\cal U}}-\sqrt{{\cal U}_0}
\right), $$ then for the $\Omega$ variable, we have
\begin{equation}
\Omega(T)=Ln \left[\left(\frac{\ell_\phi (\alpha-1)}{4(2\alpha-1)}
(T-T_0)+\sqrt{\frac{{\cal U}_0}{\ell_\phi^2}}
\right)^2-\left(\frac{\ell}{\ell_\phi}\right)^2
\right]^{\frac{2\alpha-1}{6(\alpha-1)}},
\end{equation}
and  the time transformation becomes
\begin{equation}
d\tau= \left[\left(\frac{\ell_\phi (\alpha-1)}{4(2\alpha-1)}
(T-T_0)+\sqrt{\frac{{\cal U}_0}{\ell_\phi^2}}
\right)^2-\left(\frac{\ell}{\ell_\phi}\right)^2
\right]^{\frac{1}{2(\alpha-1)}} \, dT.
\end{equation}
To obtain the solutions in the time $T$, for the anisotropic
functions $\beta_\pm(T)$, we solve the integral
\begin{eqnarray} \int e^{-3\Omega(\tau)}d\tau &=&\int
e^{-\frac{6(\alpha-1)}{2\alpha-1}\Omega(T)}dT=\int
\frac{dT}{\left(\frac{\ell_\phi (\alpha-1)}{4(2\alpha-1)}
(T-T_0)+\sqrt{\frac{{\cal U}_0}{\ell_\phi^2}}
\right)^2-\left(\frac{\ell}{\ell_\phi}\right)^2}\nonumber\\
&=& \frac{4(2\alpha-1)}{\ell (\alpha -1)} Ln \left[
\frac{\frac{\ell_\phi^2(\alpha-1)}{4(2\alpha-1))}(T-T_0)+\sqrt{\mu_0}-\ell}{\frac{\ell_\phi^2(\alpha-1)}{4(2\alpha-1))}(T-T_0)+\sqrt{\mu_0}+\ell}
\right], \label{integral}
\end{eqnarray}
then, the anisotropic functions \eqref{ani} become
\begin{equation}
 \beta_\pm (T)= \beta_\pm(T_i) \mp \frac{\ell_\pm}{3} \frac{(2\alpha-1)}{\ell
(\alpha -1)} Ln \left[
\frac{\frac{\ell_\phi^2(\alpha-1)}{4(2\alpha-1))}(T-T_0)+\sqrt{\mu_0}-\ell}{\frac{\ell_\phi^2(\alpha-1)}{4(2\alpha-1))}(T-T_0)+\sqrt{\mu_0}+\ell}
\right]. \label{ani-s}
\end{equation}
and the scalar field \eqref{scalar-field}, takes the form
\begin{equation}
\phi(T)=\phi(T_i) + \left\{
    \begin{tabular}{ll}
    $\left[\frac{(2\alpha+m)}{2\alpha}{{p^{-\frac{1}{2\alpha}}\sqrt{2\lambda}}}(T-T_i)\right]^{\frac{2\alpha}{2\alpha+m}},
    $&\quad $f(\phi)=p\phi^m, \quad m\not=-2\alpha$,\\
$    Exp\left[ {{p^{-\frac{1}{2\alpha}}\sqrt{2\lambda }}}
(T-T_i)\right],$
         &\quad $f(\phi)=p\phi^{-2\alpha}, \quad m=-2\alpha$,  \\
          $\frac{2\alpha}{m}\,Ln\left[\frac{m}{2\alpha} {{p^{-\frac{1}{2\alpha}}\sqrt{2\lambda }}}(T-T_i)\right],$ & \quad $f(\phi)=pe^{m\phi},\quad m\not=0,$\\
         ${{p^{-\frac{1}{2\alpha}}\sqrt{2\lambda }}}(T-T_i),$ & \quad $f(\phi)=p, \quad m=0,$
    \end{tabular}
    \right. \label{scalar-field-s}
\end{equation}

On the other side, the only state when the time $\tau=T$,
corresponds to the scenario, $\alpha \to \infty$, which is
calculated below.

\subsubsection{Dust scenario, $\alpha \to \infty$}
For this particular case, we have
\begin{equation}
\Omega(\tau)=Ln \left[\left(\frac{\ell_\phi\, }{8}
(\tau-\tau_0)+\sqrt{\frac{{\cal U}_0}{\ell_\phi^2}}
\right)^2-\left(\frac{\ell}{\ell_\phi}\right)^2
\right]^{\frac{1}{3}}, \label{dust-s}
\end{equation}
then the volume function becomes
\begin{equation}
V(\tau)= V_0 \left\{\left(\frac{\ell_\phi\, }{8}
(\tau-\tau_0)+\sqrt{\frac{{\cal U}_0}{\ell_\phi^2}}
\right)^2-\left(\frac{\ell}{\ell_\phi}\right)^2\right\},
\end{equation}
and the anisotropic functions are
\begin{equation}
 \beta_\pm (\tau)= \beta_\pm(\tau_0) \mp \frac{\ell_\pm}{3} \frac{2}{\ell
} Ln \left[
\frac{\frac{\ell_\phi^2}{8}(\tau-\tau_0)+\sqrt{\mu_0}-\ell}{\frac{\ell_\phi^2}{8}(\tau-\tau_0)+\sqrt{\mu_0}+\ell}
\right]. \label{ani-ss}
\end{equation}
We can see that the scalar field constant $\ell_\phi$ is huge, the
anisotropic function goes to constant, and the anisotropic model can
be isotropic one. We rewrite the corresponding solutions in the
scalar field \eqref{scalar-field}, for this scenario
\begin{equation}
\phi(\tau)=\phi(\tau_0) + \left\{
    \begin{tabular}{ll}
    $\left[{{\sqrt{2\lambda}}}(\tau-\tau_0)\right],
    $&\quad $f(\phi)=p\phi^m, \quad m\not=-2\alpha,$\\
$    Exp\left[ {{\sqrt{2\lambda }}} (\tau-\tau_0)\right],$
         &\quad $f(\phi)=, \quad m=-2\alpha,$  \\
          $\,Ln\left[{{\sqrt{2\lambda }}}(\tau-\tau_i)\right],$ & \quad $f(\phi)=pe^{2\alpha \phi},$\\
         ${{\sqrt{2\lambda }}}(\tau-\tau_0),$ & \quad $f(\phi)=p, \quad m=0.$
    \end{tabular}
    \right. \label{scalar-field-dust}
\end{equation}

\section{Quantum Regime }
The WDW equation for these models is obtained  by making the usual
substitution $ \Pi_{q^\mu}=-i \hbar \partial_{q^\mu}$ in (\ref
{hami}), and promoting the classical Hamiltonian density in the
differential operator applied to the wave function
$\Psi(\Omega,\beta_\pm,\phi)$, $\hat{\cal H}\Psi=0$; we have

\begin{equation}
\hbar^{2}
e^{-\frac{6(\alpha-1)}{2\alpha-1}\Omega}\left[-\frac{\partial^2
\Psi}{\partial \Omega^2} + \frac{\partial^2 \Psi}{\partial
\beta_+^2} +\frac{\partial^2 \Psi}{\partial \beta_-^2}
   \right]-\frac{12(2\alpha-1)}{\alpha}\,\left(\frac{2^{\alpha-1}}{\alpha{{f(\phi)}}}
\right)^{\frac{1}{2\alpha-1}}\,\hbar^{\frac{2\alpha}{2\alpha-1}}\frac{\partial^{\frac{2\alpha}{2\alpha-1}}}{\partial
\phi^{\frac{2\alpha}{2\alpha-1}}}\Psi=0. \label{q-wdw}
\end{equation}
This fractional differential equation of degree $\beta =
\frac{2\alpha}{2\alpha - 1}$, belongs to different intervals,
depending on the value of the barotropic parameter \cite{Socorro2}.
We can write this equation in terms of the  $\beta$ parameter, we
have
\begin{equation}
\hbar^{2} e^{-3(2-\beta)\Omega}\left[-\frac{\partial^2
\Psi}{\partial \Omega^2} + \frac{\partial^2 \Psi}{\partial
\beta_+^2} +\frac{\partial^2 \Psi}{\partial \beta_-^2}
   \right]-\frac{24}{\beta}\,\left(\frac{2^{\alpha-1}}{\alpha{{f(\phi)}}}
\right)^{\frac{1}{2\alpha-1}}\,\hbar^{\beta}\frac{\partial^\beta}{\partial
\phi^\beta}\Psi=0. \label{bq-wdw}
\end{equation}
For simplicity, the factor $ e^{-3(2-\beta)\Omega}$ may be the
factor ordered with $ \hat \Pi_\Omega$ and
$f^{-\frac{1}{2\alpha-1}}(\phi)$ may be the factor ordered with
$\frac{\partial ^\beta}{\partial \phi^\beta}$ in many ways, we
employ it might be called a semi-general factor ordering, which, in
this case, would order the terms  $ e^{-3(2-\beta)\Omega} \hat
\Pi^2_\Omega$ as $ - e^{-(3(2-\beta)- Q)\Omega}\,
\partial_\Omega e^{-Q\Omega}
\partial_\Omega= - e^{-3(2-\beta)\Omega}\, \partial^2_\Omega + Q\,
e^{-3(2-\beta)\Omega} \partial_\Omega, $
where $Q$ is any real constant that measures the ambiguity in the
factor ordering in the variables $ \Omega$ and its corresponding
momenta. For the other factor ordering, we make the following
calculation which, in this case, would order the terms
$\frac{g(\phi)}{f^{\frac{1}{2\alpha-1}}(\phi)}\frac{\partial
^\beta}{\partial \phi^\beta}$, where in the particular case we
choose $g(\phi)= \phi^s$, similarly to $f(\phi)$ in the classical
case, that is
\begin{equation}
f^{-\frac{1}{2\alpha-1}}(\phi)\, \phi^{-s}\frac{\partial
^{\beta/2}}{\partial \phi^{\beta/2}}\phi^s \frac{\partial
^{\beta/2}}{\partial \phi^{\beta/2}}=
f^{-\frac{1}{2\alpha-1}}(\phi)\frac{\partial^\beta}{\partial
\phi^\beta}+ f^{-\frac{1}{2\alpha-1}}(\phi)\,
\phi^{-s}\left[\frac{\partial ^{\beta/2}}{\partial
\phi^{\beta/2}}\phi^s\right] \frac{\partial ^{\beta/2}}{\partial
\phi^{\beta/2}}, \label{tra}
\end{equation}
where the Caputo fractional derivative of $\left[\frac{\partial
^{\beta/2}}{\partial \phi^{\beta/2}}\phi^s\right]$ becomes
{\cite{Podlubny}},
\begin{equation}
_0^cD_x^{\beta/2} \phi^s= \frac{\Gamma(s + 1)}{ \Gamma(s - \beta/2 +
1)} \phi^{s - \beta/2}.
\end{equation}

Thus, the equation \eqref{tra} is rewritten as
\begin{eqnarray}
f^{-\frac{1}{2\alpha-1}}(\phi)\, \phi^{-s}\frac{\partial
^{\beta/2}}{\partial \phi^{\beta/2}}\phi^s \frac{\partial
^{\beta/2}}{\partial \phi^{\beta/2}}&=&
f^{-\frac{1}{2\alpha-1}}(\phi)\frac{\partial^\beta}{\partial
\phi^\beta}+ f^{-\frac{1}{2\alpha-1}}(\phi)\,
\phi^{-s}\frac{\Gamma(s+1)}{\Gamma(s+1-\beta/2)} \phi^{s-\beta/2}
\frac{\partial ^{\beta/2}}{\partial \phi^{\beta/2}} \nonumber\\
&=&f^{-\frac{1}{2\alpha-1}}(\phi)\frac{\partial^\beta}{\partial
\phi^\beta}+ f^{-\frac{1}{2\alpha-1}}(\phi)\,
\frac{\Gamma(s+1)}{\Gamma(s+1- \beta/2)} \phi^{-\beta/2}
\frac{\partial ^{\beta/2}}{\partial \phi^{\beta/2}}
\end{eqnarray}

Assuming this factor ordering for the Wheeler-DeWitt equation, we
get
\begin{eqnarray}
&& e^{-3(2-\beta)\Omega}\left[-\frac{\partial^2 \Psi}{\partial
\Omega^2} + Q \frac{\partial \Psi}{\partial \Omega}+
\frac{\partial^2 \Psi}{\partial \beta_+^2} +\frac{\partial^2
\Psi}{\partial \beta_-^2}
   \right] \nonumber\\
&& -\frac{24}{\beta}\,\left(\frac{2^{\alpha-1}}{\alpha{{f(\phi)}}}
\right)^{\frac{1}{2\alpha-1}}\,\hbar^{\beta}\frac{\partial^\beta}{\partial
\phi^\beta}\Psi-\frac{24}{\beta}\,\left(\frac{2^{\alpha-1}}{\alpha{{f(\phi)}}}
\right)^{\frac{1}{2\alpha-1}}\,\hbar^{\beta}\phi^{-s}\frac{\Gamma(s+1)}{\Gamma(s+1-\mu)}\frac{\partial^\frac{\beta}{2}}{\partial
\phi^\frac{\beta}{2}}\Psi=0. \label{bqq-wdw}
\end{eqnarray}
Using the ansatz for the wave function
$\Psi(\Omega,\beta_+,\beta_-,\phi)={\cal A}(\Omega){\cal
B_+}(\beta_+){\cal B_-}(\beta_-){\cal C}(\phi)$, we obtain the
following differential equations on the corresponding variables
\begin{eqnarray}
\frac{d^2{\cal A}}{d\Omega^2}- Q \frac{d {\cal
A}}{d\Omega}-\left[\pm \left( \frac{\rho}{\hbar}\right)^2\,
e^{3(2-\beta)\Omega}+\rho_1^2 \right] {\cal A}&=&0,
\label{om}\\
\frac{d^2{\cal B_+}}{d\beta_+^2}-\rho_2^2{\cal B_+}&=&0, \label{b+}\\
\frac{d^2{\cal B_-}}{d\beta_-^2}-\rho_3^2{\cal B_-}&=&0,
\qquad \rho_3^2=\rho_2^2-\rho_1^2\label{b-}\\
 \phi^\gamma \frac{d^{2\gamma}{\cal C}}{d\phi^{2\gamma}}+\frac{\Gamma(s+1)}{\Gamma(s+1-\gamma)}\frac{d^{\gamma}
{\cal C}}{d\phi^{\gamma}}\pm \left(\frac{p_\phi
\alpha}{2^{\alpha-1}} \right)^\frac{1}{2\alpha-1} \frac{\gamma
\rho^2}{12\hbar^{2\gamma}} {\cal C}&=&0, \quad \beta=2\gamma, \quad
0<\gamma \leq 1. \label{phi-1}
\end{eqnarray}
We can see that the fractional differential equation \eqref{phi-1}
has variable coefficients, so to solve it we can use the fractional
power series \cite{El-Ajou, Rida}, also $f(\phi)=p_\phi \phi^m$,
with $p_\phi$ a constant and choosing $\frac{m}{2\alpha-1}=-\gamma$,
as particular case,   which implies that the parameter $m=-\alpha$
in sense that $\gamma$ parameter is in accordance with its original
definition (see equation (\ref{phi-1}) and definition of the $\beta$
parameter, or the equation in the text after equation
(\ref{nueva})).

Following the book of Polyanin \cite{polyanin} (page 179.10), we
find the solution for the first equation, considering different
values in the factor ordering parameter, (we take the corresponding
sign minus in the constant $\rho^2$)

\begin{equation}
\mathcal{A}(\Omega) =
e^{\frac{Q\Omega}{2}}\biggl[C_{1}K_{\nu}\biggl(\frac{\rho}{3\hbar(1
- \gamma)}e^{3(1 - \gamma)\Omega}\biggr) +
C_{2}I_{\nu}\biggl(\frac{\rho}{3\hbar(1 - \gamma)}e^{3(1 -
\gamma)\Omega}\biggr)\biggr], \label{nueva}
\end{equation}
where $K_{\nu}(z)$ and $I_{\nu}(z)$ are the modified Bessel
functions,
 and order $\nu = \frac{\sqrt{Q^2+4\rho_1^2}}{6(1-\gamma)}$ with
$\gamma = \frac{\beta}{2}=\frac{\alpha}{2\alpha -1}$ the new order
in the fractional derivative. However, for physical conditions we
will only take the modified Bessel $K_{\nu}$.

The corresponding quantum solution for equations \eqref{b+} and
\eqref{b-} are
\begin{eqnarray}
\mathcal{B}_{+} = a_0 e^{\rho_2 \beta_+} + a_1 e^{-\rho_2
\beta_+},\label{solb+}\\
\mathcal{B}_{-} = b_0 e^{\rho_3 \beta_-} + b_1 e^{-\rho_3
\beta_-}.\label{solb-}
\end{eqnarray}
with $\rho^2$ and  $\rho_i^2$ are  separation constants.

\begin{figure}[!htbp]
    \centering
        \includegraphics[width=0.4\textwidth]{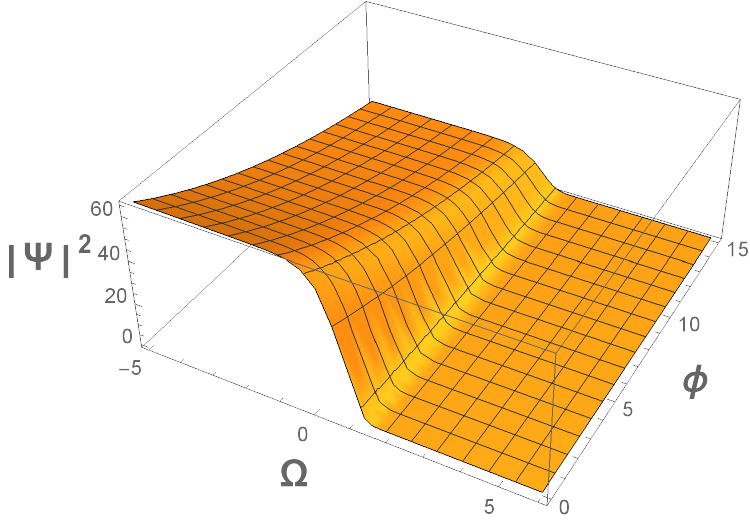}
        \includegraphics[width=0.4\textwidth]{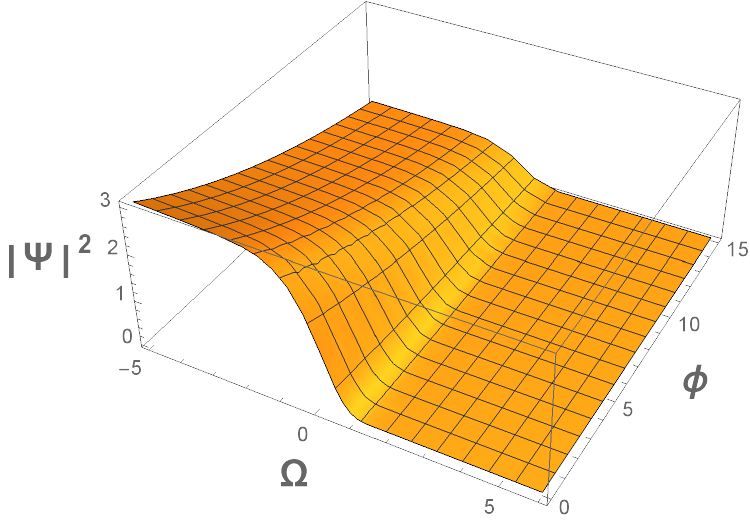}
\caption{Probability density \eqref{psi0} of the universe dominated
by Dust era, in this stage of the universe shows that the
probability density has a decay, both in $\Omega$ and $\phi$. Both
are plotting with $\rho=0.5$, $\gamma=\frac{1}{2}$, $\rho_1=0.1$; to
the left figure we take $Q = 6$, $\nu = 2.00111$, and right figure
$Q = 2$, $\nu = 0.669992$. }\label{todefine}
\end{figure}

The solution of the equation \eqref{phi-1} with positive sign and
$f(\phi)=p_\phi=constant$, with zero factor ordering, may be
obtained by applying direct and inverse Laplace transform
\cite{Rosales2,Socorro2}, providing

\begin{equation}
 {\cal C}_+ (\phi, \gamma) =
 \mathbb{E}_{2\gamma}\left(- z^2\right),
 \qquad z={\left(\frac{p_\phi \alpha}{2^{\alpha-1}}
\right)^{\frac{1}{2(2\alpha-)}}} \frac{\sqrt{\gamma} \rho
}{2\sqrt{3}\hbar^{\gamma}} \phi^\gamma,\qquad 0<\gamma \leq 1,
\label{R}
 \end{equation}
where $\mathbb{E}_{2\gamma}$  is the Mittag-Leffler function
(\ref{C5}), then, the probability density of the wave function for
this particular case becomes
\begin{equation} |\Psi|^2=\psi_0^2 e^{Q\Omega \pm 2\rho_2 \beta_+ \pm
2\rho_3\beta_-}\,K_\nu^2\left[\frac{\rho}{3\hbar(1-\gamma)}
e^{3(1-\gamma)\Omega} \right]\, \mathbb{E}^2_{2\gamma}(-z^2),
\label{psi0}
\end{equation}
and its corresponding plot for two values in the ordering parameter
Q, as is shown in figure \ref{todefine}.

Table II shows the differential equations obtained from
\eqref{phi-1}, depending on the values of $\omega_X$, $\alpha$ and
$\gamma$

\begin{table}[!htbp]
  \centering
  \begin{tabular}{|c|c|c|c|}
\hline
  $\omega_{X}$ & $\alpha$ & $\gamma$ & Fractionary equation \\
\hline\hline 1 & 1 & 1 & $ \phi \frac{d^{2}{\cal
C}}{d\phi^{2}}+\frac{\Gamma(s+1)}{\Gamma(s)}\frac{d {\cal
C}}{d\phi}\pm p_\phi \frac{ \mu^2}{24\hbar^{2}}
{\cal C}=0 $ \\
\hline $\frac{1}{3}$  & 2 & $\frac{2}{3}$ &$\phi^{\frac{2}{3}}
\frac{d^{\frac{4}{3}}{\cal
C}}{d\phi^{\frac{4}{3}}}+\frac{\Gamma(s+1)}{\Gamma(s+\frac{1}{3})}\frac{d^{\frac{2}{3}}
{\cal C}}{d\phi^{\frac{2}{3}}}\pm (p_\phi)^\frac{1}{3} \frac
{\mu^2}{18\hbar^{\frac{4}{3}}} {\cal C}=0 $ \\\hline $\rightarrow 0$
& $\rightarrow \infty $ & $\frac{1}{2}$  & $\phi^{\frac{1}{2}}
\frac{d{\cal
C}}{d\phi}+\frac{\Gamma(s+1)}{\Gamma(s+\frac{1}{2})}\frac{d^{\frac{1}{2}}
{\cal C}}{d\phi^{\frac{1}{2}}}\pm

\frac{ \mu^2}{24\hbar} {\cal C}=0 $ \\\hline $-1$ & 0 & 0 &
$without\,\, equation$ \\\hline $-\frac{1}{3}$ & -1 & $\frac{1}{3}$
& $\phi^{\frac{1}{3}} \frac{d^{\frac{2}{3}}{\cal
C}}{d\phi^{\frac{2}{3}}}+\frac{\Gamma(s+1)}{\Gamma(s+\frac{2}{3})}\frac{d^{\frac{1}{3}}
{\cal C}}{d\phi^{\frac{1}{3}}}\pm \left(\frac{p_\phi
\alpha}{2^{\alpha-1}} \right)^\frac{1}{2\alpha-1} \frac{
\mu^2}{36\hbar^{\frac{2}{3}}} {\cal C}=0$\\\hline $-\frac{2}{3}$  &
$-\frac{1}{4}$ & $\frac{1}{6}$ & $ \phi^{\frac{1}{6}}
\frac{d^{\frac{1}{3}}{\cal
C}}{d\phi^{\frac{1}{3}}}+\frac{\Gamma(s+1)}{\Gamma(s+\frac{5}{6})}\frac{d^{\frac{1}{6}}
{\cal C}}{d\phi^{\frac{1}{6}}}\pm \left(\frac{p_\phi
\alpha}{2^{\alpha-1}} \right)^\frac{1}{2\alpha-1} \frac{
\mu^2}{72\hbar^{\frac{1}{3}}}
{\cal C}=0$ \\
\hline
  \end{tabular}
  \caption{Fractionary equation in the field $\phi$ according to the barotropic parameter $\omega_{X}$.}
  \label{fra-equation}
\end{table}

\subsection{Solution to FDE associated with the different state evolutions}

We write the fractional differential equation \eqref{phi-1} as
follows,
\begin{equation}
\phi^{\gamma}\frac{d^{2\gamma}{\mathcal{C}}}{d\phi^{2\gamma}} + A
\frac{d^{\gamma}\mathcal{C}}{d\phi^{\gamma}} + B\,\mathcal{C}=0,
\quad 0<\gamma \leq 1, \label{DFDE}
\end{equation}
where we have made the simplifications $A=
\frac{\Gamma(s+1)}{\Gamma(s+1-\gamma)}$ and $B_{(\alpha, \gamma)}=
\pm \left(\frac{p_\phi \alpha}{2^{\alpha-1}}
\right)^\frac{1}{2\alpha-1} \frac{\gamma \mu^2}{12\hbar^{2\gamma}}$.
The last linear fractional differential equation \eqref{DFDE}, will
be solved using the fractional power series \cite{El-Ajou, Rida}

\begin{equation}
\mathcal{C} = \sum^{\infty}_{n=0} a_{n}\,\phi^{n\gamma}.
\label{serD}
\end{equation}
Then, the fractional derivatives are

\begin{subequations}
\label{derser}
  \begin{align}
    \frac{d^{\gamma}\mathcal{C}}{d^{\gamma}\phi} & = \sum_{n=1}^{\infty}a_{n}\,\frac{\Gamma[n\gamma + 1]}{\Gamma[(n - 1)\gamma + 1]}\,\phi^{(n - 1)\gamma}, \\
    \frac{d^{2\gamma}\mathcal{C}}{d^{2\gamma}\phi} & = \sum_{n=2}^{\infty}a_{n}\,\frac{\Gamma[n\gamma + 1]\Gamma[(n - 1)\gamma + 1]}{\Gamma[(n - 1)\gamma + 1]\Gamma[(n - 2)\gamma + 1]}\,\phi^{(n - 2)\gamma}.
  \end{align}
\end{subequations}
Replacing the expressions \eqref{derser} into \eqref{DFDE}, we get

\begin{equation}
\sum_{n=2}^{\infty}a_{n}\,\frac{\Gamma[n\gamma + 1]}{\Gamma[(n -
2)\gamma + 1]}\,\phi^{(n - 1)\gamma} +
A\sum_{n=1}^{\infty}a_{n}\,\frac{\Gamma[n\gamma + 1]}{\Gamma[(n -
1)\gamma + 1]}\,\phi^{(n - 1)\gamma} + B_{(\alpha, \gamma)}
\sum^{\infty}_{n=0} a_{n}\,\phi^{n\gamma} = 0. \label{subser}
\end{equation}
Now, taking $\ell = n - 1$ into the first and second terms, and $n =
\ell$ into the third term of \eqref{subser}, we have

\begin{equation}
\sum_{\ell=1}^{\infty}a_{\ell + 1}\,\frac{\Gamma[(\ell + 1)\gamma +
1]}{\Gamma[(\ell - 1)\gamma + 1]}\,\phi^{\ell \gamma} + A\sum_{\ell
= 0}^{\infty}a_{ \ell + 1}\,\frac{\Gamma[(\ell + 1)\gamma +
1]}{\Gamma[\ell \gamma + 1]}\,\phi^{\ell \gamma} + B_{(\alpha,
\gamma)} \sum^{\infty}_{\ell=0} a_{\ell}\,\phi^{\ell \gamma} = 0.
\end{equation}
Shifting one place in the second and third summations, we have

\begin{align}
\sum_{\ell=1}^{\infty}a_{\ell + 1}\,\frac{\Gamma[(\ell + 1)\gamma +
1]}{\Gamma[(\ell - 1)\gamma + 1]}\,\phi^{\ell \gamma} + &
A\,\frac{\Gamma[\gamma + 1]}{\Gamma[1]}a_{1}
+ A\sum_{\ell = 1}^{\infty}a_{ \ell + 1}\,\frac{\Gamma[(\ell + 1)\gamma + 1]}{\Gamma[\ell \gamma + 1]}\,\phi^{\ell \gamma} \nonumber\\
& + B_{(\alpha, \gamma)}a_{0} + B_{(\alpha, \gamma)}
\sum^{\infty}_{\ell=1} a_{\ell}\,\phi^{\ell \gamma} = 0.
\label{derser1}
\end{align}
From last expression \eqref{derser1}, we get $(s\not=0)$

\begin{equation}
a_{1} = - \frac{B_{(\alpha,\gamma)}}{ A \Gamma(\gamma + 1)} = -
\frac{ \Gamma[s+1-\gamma] B_{(\alpha,
\gamma)}}{\Gamma[s+1]\,\Gamma[\gamma + 1]}a_{0}, \label{firsterm}
\end{equation}
and the recurrence relationships between the parameters $a_\ell$, is
\begin{equation}
a_{\ell + 1} = -\frac{\Gamma[s+1-\gamma]B_{(\alpha,
\gamma)}\,\Gamma[\ell \gamma + 1]\,\Gamma[(\ell - 1)\gamma + 1]\,
}{\Gamma[(\ell + 1)\gamma +
1]\biggl\{\Gamma[s+1-\gamma]\,\Gamma[\ell \gamma + 1] +
\Gamma[s+1]\,\Gamma[(\ell - 1)\gamma + 1]\biggr\}}\,a_{\ell}, \qquad
\forall\, \ell\geq 1.
\end{equation}
Some terms of this relation are

\begin{align}
 a_{1} & = - \frac{ \Gamma[s+1-\gamma]B_{(\alpha, \gamma)}
}{\Gamma[s+1]\,\Gamma[\gamma +
1]}a_{0},\nonumber\\
a_2 & = \frac{\left(\Gamma[s+1-\gamma]B_{(\alpha,
\gamma)}\right)^2}{\Gamma[s+1]\Gamma[2\gamma+1]\left(\Gamma[s+1]+\Gamma[s+1-\gamma]\Gamma[\gamma+1]
\right)}
a_0,\nonumber\\
a_{3} & = -\frac{\Gamma[\gamma+
1]}{\Gamma[s+1]\Gamma[3\gamma+1]\left(\Gamma[s+1-\gamma]\Gamma[2\gamma+1]+\Gamma[s+1]\Gamma[\gamma+1]
\right)}\times \nonumber\\
\label{acoeff} & \qquad \qquad
\Biggl(\frac{\left(\Gamma[s+1-\gamma]B_{(\alpha,
\gamma)}\right)^3}{\left(\Gamma[s+1]+\Gamma[s+1-\gamma]\Gamma[\gamma+1] \right)}\Biggr) a_{0}, \\
a_4 & = \frac{\Gamma[\gamma+1]\Gamma[2\gamma+1]}{\Gamma[s+1]\Gamma[4\gamma+1]\left(\Gamma[s+1-\gamma]\Gamma[3\gamma+1] + \Gamma[s+1]\Gamma[2\gamma+1]\right)}\times \nonumber\\
& \qquad\Biggl( \frac{(\Gamma[s + 1-
\gamma]\,B_{(\alpha,\gamma)})^{4}}{\left(\Gamma[s+1\gamma]\Gamma[2\gamma+1]+\Gamma[s+1]\Gamma[\gamma+1]
\right)\left(\Gamma[s+1]+\Gamma[s+1-\gamma]\Gamma[\gamma+1] \right)}\Biggr)a_{0}, \nonumber \\
\vdots  \nonumber
\end{align}

Then, the solution of the fractional equation \eqref{DFDE} has the
form
\begin{eqnarray}
\mathcal{C}_{s,\alpha,\gamma} &=&  a_0\left[1 -
\frac{\Gamma[s+1-\gamma]B_{(\alpha,
\gamma)}}{\Gamma[s+1]\,\Gamma[\gamma + 1]}\phi^\gamma \right. \label{serDE}\\
&& \left. + \frac{\left( \Gamma[s+1-\gamma]B_{(\alpha,
\gamma)}\right)^2}{\Gamma[s+1]\Gamma[2\gamma+1]\left(\Gamma[s+1]+\Gamma[s+1-\gamma]\Gamma[\gamma+1]
\right)} \phi^{2\gamma} + \ldots \right]. \nonumber
\end{eqnarray}

For the Dust-like scenario (see table \ref{fra-equation}), $\alpha
\to \infty$ and $\gamma=\frac{1}{2}$, then  $B_{(\infty,
\frac{1}{2})}=\frac{\mu^2}{24\hbar}$. The solution associated with
this fractional differential equation is given in fractional series
form by,

\begin{eqnarray}
\mathcal{C}_{s,\to \infty,\frac{1}{2}} &=& a_{0}\left[1 -
\frac{2\Gamma[s+\frac{1}{2}]B_{(\infty,
\frac{1}{2})}}{\Gamma[s+1]\sqrt{\pi}}\,\phi^{\frac{1}{2}} +
\frac{2\left(\Gamma[s+\frac{1}{2}]B_{(\infty,
\frac{1}{2})}\right)^{2}}{\Gamma[s+1](2\Gamma[s+1] +
\Gamma[s+\frac{1}{2}]\sqrt{\pi})}\,\phi \right. \label{FDEDust}\\
&&\left. - \frac{2\left(2\Gamma[s+\frac{1}{2}]B_{(\infty,
\frac{1}{2})}\right)^{3}}{3\sqrt{\pi}\Gamma[s+1](2\Gamma[s+\frac{1}{2}]
+ \Gamma[s+1]\sqrt{\pi})(2\Gamma[s+1] +
\Gamma[s+\frac{1}{2}]\sqrt{\pi})}\,\phi^{\frac{3}{2}} +
\cdots\right],
  \nonumber
\end{eqnarray}
where we employed $\Gamma[\frac{3}{2}]=\frac{\sqrt{\pi}}{2}$ and
$\Gamma[\frac{5}{2}]=\frac{3\sqrt{\pi}}{4}$.

For the radiation stage, $\alpha=2$ and $\gamma=\frac{2}{3}$, then
$B_{(2,\frac{2}{3})}=\frac{\sqrt[3]{p_\phi}\mu^2}{18\hbar^{\frac{4}{3}}}$,
the solution for the fractional differential equation is

\begin{eqnarray}
\mathcal{C}_{s, 2,\frac{2}{3}} &=& a_{0}\left[1 -
\frac{3\Gamma[s+\frac{1}{3}]B_{(2,
\frac{2}{3})}}{2\Gamma[s+1]\Gamma[\frac{2}{3}]}\,\phi^{\frac{2}{3}}
+ \frac{\left(3\Gamma[s+\frac{1}{3}]B_{(2,
\frac{2}{3})}\right)^{2}}{4\Gamma[s+1]\Gamma[\frac{4}{3}]\left(
3\Gamma[s+1]
+2\Gamma[s+\frac{1}{3}]\Gamma[\frac{2}{3}]\right)}\,\phi^{\frac{4}{3}} \right. \label{FDErad}\\
&&\left. -
\frac{\frac{2}{3}\Gamma[\frac{2}{3}]\left(3\Gamma[s+\frac{1}{3}]B_{(2,
\frac{2}{3})}\right)^{3}}{2\Gamma[s+1]\left(2\Gamma[s+\frac{1}{3}]\Gamma[\frac{4}{3}]
+ \Gamma[s+1]\Gamma[\frac{2}{3}]\right)\left(3\Gamma[s+1] +
2\Gamma[s+\frac{1}{3}]\Gamma[\frac{2}{3}]\right)}\,\phi^{\frac{8}{3}}
+ \cdots\right], \nonumber
 \end{eqnarray}
with $\Gamma[\frac{2}{3}]=1.35412$ and
$\Gamma[\frac{4}{3}]=0.89298$.

\begin{figure}[!htbp]
    \centering
            \includegraphics[width=0.4\textwidth]{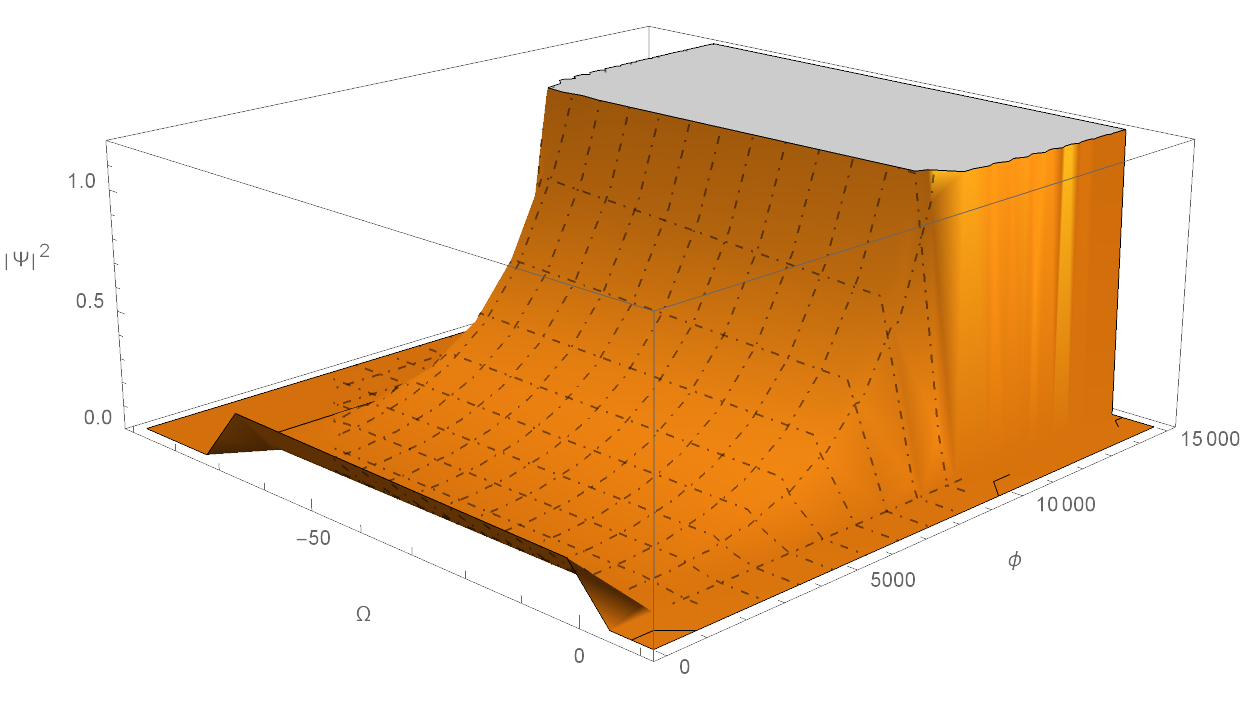}
\includegraphics[width=0.4\textwidth]{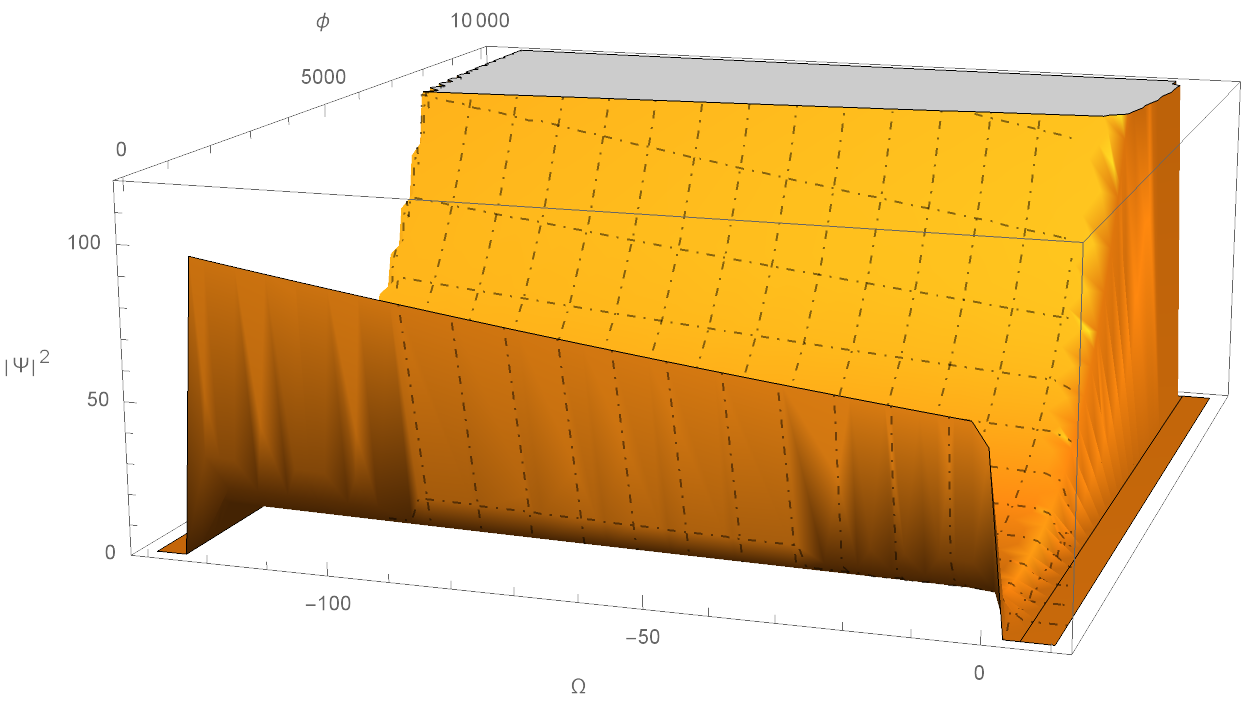}
\includegraphics[width=0.4\textwidth]{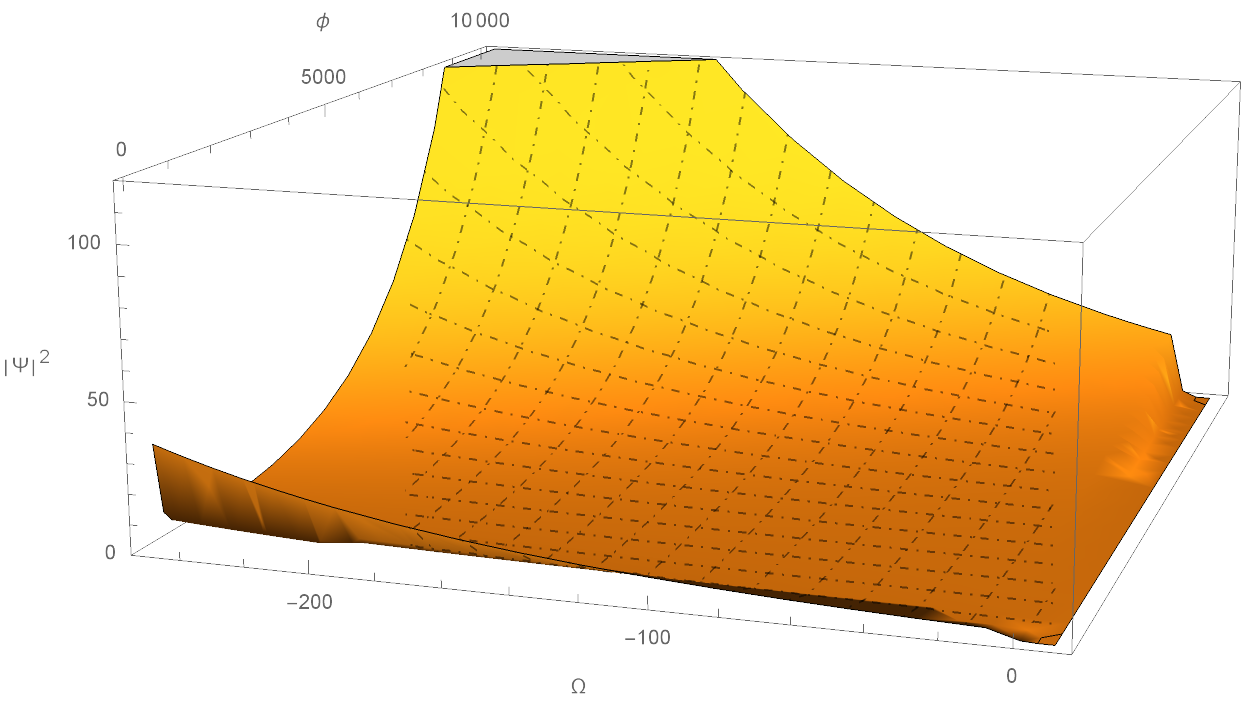}
\includegraphics[width=0.4\textwidth]{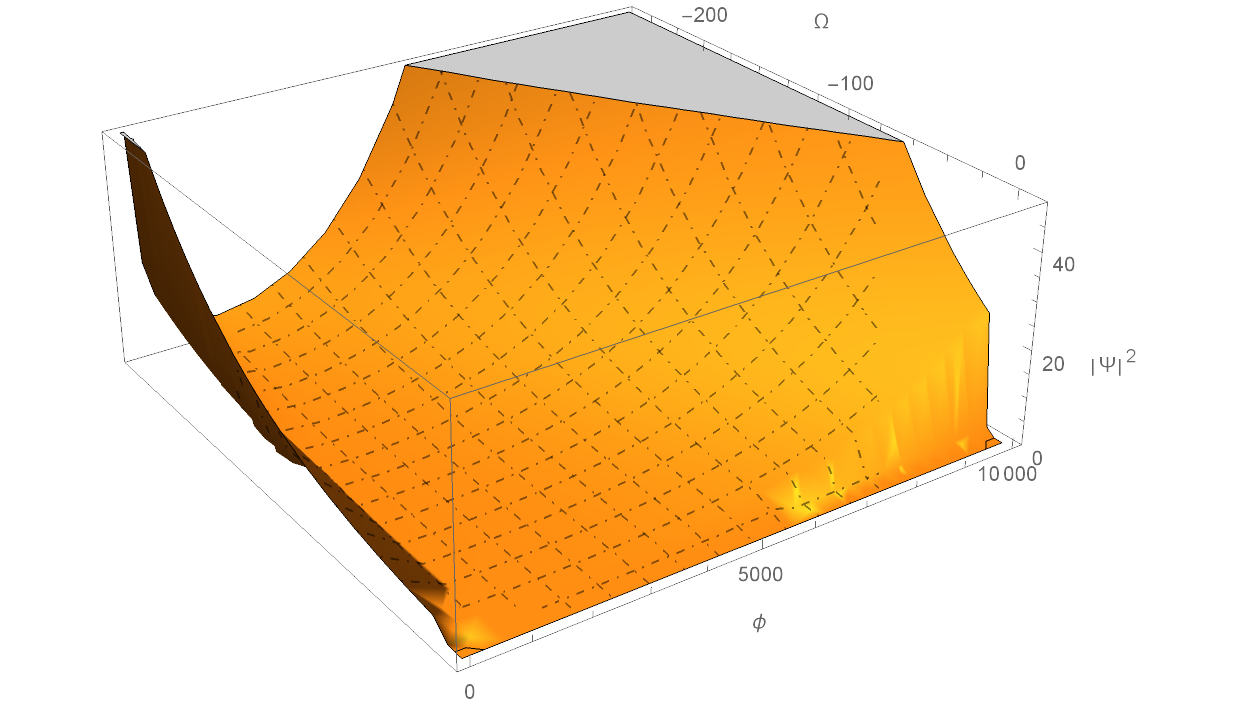}
    \caption{Probability density \eqref{simplewave} of the universe dominated by Dust era, in this stage of the universe shows
    that the probability density has a decay in  $\Omega$ and exhibits considerable growth for certain values of  $\phi$.
    The plots have the parameters $\rho=0.5$, $\gamma=\frac{1}{2}$, $\rho_1=0.1$.  The Q and $\nu$ parameters taken the values,
    $Q = 10,6,2,1.6 $, $\nu = 3.334$,2.00111,0.0669992,0.537489, respectively,}\label{probden}
\end{figure}

We are going to present the graphical behaviour of the wave function
for the Dust-like case, in which the solution is constrained to the
variables $\Omega$ and $\phi$. This means we are going to shrink the
directions $\beta_{+}$ and $\beta_{-}$. So, the wave function takes
the form
\begin{align}
\Psi(\Omega,  \beta_{+},\beta_{-},\phi) & =
e^{\frac{Q\Omega}{2}}\biggl[C_{1}K_{\nu}\biggl(\frac{\rho}{3\hbar(1
- \gamma)}e^{3(1 - \gamma)\Omega}\biggr)\biggr] \times \nonumber \\
& \Biggl[1 - a_1\phi^{\frac{1}{2}} + a_2 \phi  - a_3
\phi^{\frac{3}{2}} + a_4 \phi^2 - a_5 \phi^{\frac{5}{2}} +
  a_6 \phi^3 - a_7 \phi^{\frac{7}{2}} + a_8 \phi^4+ \ldots\Biggr].
\label{wavetot}
\end{align}
By restraining ourselves to the values of $\nu = \frac{\sqrt{Q^2 + 4
\rho_1^2}}{6 (1 - \gamma))}$, $Q= 10,6,2,1.6$, $\rho=\frac{1}{2}$,
$\gamma = \frac{1}{2}$, $s=1$, $\rho_1=0.1$ and $C_{1} = 1$, we see
that the wave function (\eqref{wavetot}) can be rewritten as
\begin{eqnarray}
\Psi^2(\Omega,\phi)& =&
e^{Q\Omega}\,K^2_{\nu}\left(\frac{2}{3}e^{\frac{3\Omega}{2}}\right)
\left[1 - \frac{1}{24}\phi^{1/2} + \frac{\sqrt{\pi}}{576(4+\pi)}\phi
- \frac{2\sqrt{\pi}}{41472(4 + \pi)}\phi^{3/2}  \right.\nonumber\\
&& \left. +\frac{\pi^{\frac{3}{2}}}{663552(8+3\pi)(4+\pi)} \phi^2
- \frac{\pi^{\frac{3}{2}}}{69672960(8+3\pi)(4+\pi)} \phi^{\frac{5}{2}} \right. \nonumber\\
&& \left.+ \frac{\pi^{\frac{5}{2}}}{334430208(8+3\pi)(4+\pi)(32+15\pi)} \phi^3 \right. \nonumber\\
&& \left.- \frac{\pi^{\frac{5}{2}}}{45649723392(8+3\pi)(4+\pi)(32+15\pi)} \phi^{\frac{7}{2}} \right. \nonumber\\
&&\left.+
\frac{5\pi^{\frac{7}{2}}}{208684449792(8+3\pi)(4+\pi)(32+15\pi)(192+105\pi)}
\phi^4 \right]^2\,, \label{simplewave}
\end{eqnarray}
where we have taken the cut-off order in $\phi^4$ and the a's
parameter are read from \eqref{acoeff}. In the following, we present
some plots of the probability density of the wave function,
including the factor ordering parameter Q and particular values in
the parameter $\rho$, $\rho_1$ and the particular value to the
ordering parameter s=1. In all of them we observe that for any value
of Q, the probability density decays with respect to the scale
factor, but has a different evolution in the scalar field. For small
Q's, the quantum universe has considerable existence in the
evolution with respect to the scale factor and then decays. On the
other hand, for large Q's, this interval is small. What we can say
about the evolution of the scalar field is that at small Q's, the
scalar field appears faster than for large Q's, which enters late,
but has existed forever (see figure \ref{probden}).

\section{Conclusions}
Unlike the previous work \cite{Socorro2}, in the present paper we
employed a barotropic equation with perfect fluid for
energy-momentum tensor in the K-essence scalar field into the
Lagrangian and Hamiltonian formalism, obtaining the momentum of a
scalar field with fractional numbers, while the momentum of the
scale factor appears in the usual way. We obtained the classical
solutions for different scenarios in the universe, employing
different times $(t,T(\tau),\tau)$. In the quantum scheme, we
include the factor ordering problem, and we find a fractional
differential equation for the scalar field with variable
coefficients, which was solved using the fractional series
expansion. With this in mind, we visualize two alternatives in our
analysis; the first one is within the traditional expectation over
the behaviour of the probability density, that the best candidates
for quantum solutions are those that have a damping behaviour with
respect to the scale factor, appearing in all scenarios under our
study, without saying anything about the scalar field. The other
scenario is when we keep the scale factor, and we consider the
values of the scalar field as significant in the quantum regime,
appearing in various scenarios in the behaviour of the universe;
mainly in those where the universe has a huge behaviour, for
example, in the actual epoch, where the scalar field appears as
background.

In other words, the interpretation of the probability density of the
unnormalized wave function is given, when we demand that $\Psi$ does
not diverge when the scale factor A (or $\Omega$) goes to infinity,
and the scalar field is arbitrary.  However, the  evolution with the
scalar field is important in this class of theory and others, as it
appears in some stages of the evolution of our Universe.

In the reference \cite{Micolta-2023}, the gravitational action
integral is altered by hand, leading to a modified Friedmann-type
equation. They employed the dynamical system approach in order to
find the balance points providing a range for the order of the
fractional derivatives in their investigation of the cosmological
universe, and they do mention that it can be confirmed that the
solutions isotropize at a late time. In our approach, this occurs
when the $\ell_\phi$ is huge (see equation \eqref{ani-ss}); due to
that, the anisotropic parameters become constants. On the other
hand, as we use the volume of the universe $V=ABC=e^{3\Omega(\tau)}$
on any time scale, it depends only on the $\Omega$ function  and not
on the anisotropic parameters $\beta_\pm$. Not so, the scale factors
$A$, $B$ and $C$, which do depend on these. It remains to study
whether the fractional derivatives alter the gravitational part and
how the universe's singularity can be avoided because, from our
approach, this part needs to be revised since the gravitational part
is not altered from the point of view of the equations of motion. It
is modified in the scalar field part.

It would be interesting to extend the Bohm-type semiclassical
formalism in this context, which we will do in future work. Much
work has been made in this direction
\cite{paulo2012,Priscila2013,1511.08382,omar2017,1710.08666,2012.10814,2103.00802},
where the quantum potential emerges as the imaginary part in the
Bohm formalism, appearing as a constraint equation. In this sense,
in the reference \cite{Jalalzadeh-2023} appears an approach based on
the semiclassical limit of fractionary quantum cosmology using the
Riesz derivative. It would be interesting to continue under our
focus, where the corresponding Friedmann equation and the Hubble
parameter depend on the Levy's fractional parameter, which is
associated with the concept of L\'evy path in the corresponding
quantum cosmology.

We briefly illustrate the main results of this work.

\begin{enumerate}
\item{} Using the K-essence formalism in a general way, applied to anisotropic Bianchi type I cosmological model, we found the Hamiltonian density in the scalar field momenta raised to power with non-integers, which produces in the quantum scheme a
fractional differential equation in a natural way. We include the
factor ordering problem in both variables $(\Omega,\phi)$ and its
momenta $(\Pi_\Omega, \Pi_\phi)$, with the order
$\beta=\frac{2\alpha}{2 \alpha-1}$, where $\alpha \in (-1,\infty)$,
and it was solved in a general way, we include two particular
scenarios of our Universe.

\item{} We found the solution in the classical scheme employing two gauges, $N=24e^{
3\Omega}$, for two forms of the function $f(\phi)$ in the time $t$;
however, when we let the Lagrange multiplier N, we need to employ a
transformed time $T(\tau)$ for solving the classical equation, and
only in the dust era, we recover the gauge time $\tau$.

\item{} In the quantum regime, when we include the factor ordering problem, the fractional differential equation in the scalar field appears with variable coefficients, and it was necessary to use the fractional series expansion to solve it in a general way.

\item{} In one of our analysis presented on the probability density, we consider the
 values of the scalar field as significant in the quantum regime, appearing in various scenarios in
the behaviour of the universe; mainly in those where the universe
has a huge behaviour, for example, in the actual epoch, where the
scalar field appears as a background, the quantum regime appears
with big values, but it presents a moderate development in other
scenarios with different ordering parameters Q and s.
\end{enumerate}


\acknowledgments{ \noindent J.S. was partially supported by PROMEP
grants UGTO-CA-3. Authors were partially supported SNI-CONACyT. J.
Rosales is supported by PROMEP grants UGTO-CA-20 nonlinear photonics
and the Department of Electrical Engineering. This work is part of
the collaboration within the Instituto Avanzado de Cosmolog\'{\i}a
and Red PROMEP: Gravitation and Mathematical Physics under project
{\it Quantum aspects of gravity in cosmological models,
phenomenology and geometry of space-time.} L.T.S. is supported by
Secretaria de Investigaci\'on y Posgrado del Instituto Polit\'ecnico
Nacional, grant SIP20211444. Many calculations were done by Symbolic
Program REDUCE 3.8.}.

 \end{document}